\documentclass[a4paper]{article}
\usepackage{amssymb}
\usepackage{latexsym}
\newcommand{\Zz}{{\mathbb Z}}

\newcommand{\Rr}{{\mathbb R}}

\def\be{\begin{equation}}
\def\ee{\end{equation}}
\def\bea{\begin{eqnarray}}
\def\eea{\end{eqnarray}}

\def\d{{\,\rm d}}

\def\0{{\bf 0}}

\def\k{{\bf k}}
\def\e{{\bf e}}

\def\q{{\bf q}}
\def\n{{\bf n}}
\def\r{{\bf r}}
\def\p{{\bf p}}

\def\x{{\bf x}}
\def\y{{\bf y}}
\def\K{{\bf K}}

\def\R{{\bf R}}

\def\h2m{\frac{\hbar^2}{2m}}
\def\p0{{P_{\beta H^0_N}}}

\begin{document}
\title{
\large\bf A possible mechanism of concurring diagonal and off-diagonal long-range order for soft interactions}
\author{Andr\'as S\"ut\H o\\
Research Institute for Solid State Physics and Optics\\ Hungarian Academy of Sciences, P. O. Box 49, H-1525 Budapest, Hungary\\ E-mail: suto@szfki.hu}
\thispagestyle{empty}
\date{}
\maketitle
\begin{abstract}
\noindent
This paper is a contribution to the theory of coherent crystals. We present arguments claiming that negative minima in the Fourier transform of a soft pair interaction may give rise to the coexistence of diagonal and off-diagonal long-range order at high densities, and that this coexistence may be detectable due to a periodicity seen on the off-diagonal part of the one-body reduced density matrix, without breaking translation invariance. As an illustration, we study the ground state of a homogenous system of bosons in continuous space, from the interaction retaining only the Fourier modes $v(\k)$ belonging to a single nonzero wave number $|\k|=q$. The result is a mean-field model. We prove that for $v(\k)>0$ the ground state is asymptotically fully Bose-condensed, while for $v(\k)<0$ at densities exceeding a multiple of $\hbar^2 q^2/2m|v(\k)|$ it exhibits both Bose-Einstein condensation and diagonal long-range order, and the latter can be seen on both the one- and the two-body density matrix.

\vspace{2mm}

\vspace{2mm}
\noindent

\end{abstract}

\section{Introduction}
Ever since the clarification by Penrose and Onsager \cite{PO} of the notion of Bose-Einstein condensation (BEC) for interacting bosons, an important question has been the accommodation of a diagonal long-range order (DLRO) with the off-diagonal one (ODLRO), characterizing BEC. This question has gained a new interest with the experimental discovery of nonclassical rotational inertia in solid helium \cite{KC}, its being a possible signature of coexistence of the two types of order \cite{Leg}. Penrose and Onsager defined BEC as the non-decay of $\lambda_N/N$ with $N$ going to infinity, where $N$ is the number of particles and $\lambda_N$ is the largest eigenvalue of the one-body reduced density matrix. If the Hamiltonian is translation invariant, as in the case when it contains only a shift-invariant interaction and no external field and is defined with periodic boundary conditions on a cube $\Lambda$, a Perron-Frobenius argument shows that $\lambda_N=\langle a_\0^*a_\0\rangle$, the expectation value of the occupation number of the zero-momentum one-particle state. Expressing $a_\0^{(*)}$ in terms of the field operators $a^{(*)}(\x)$, one obtains $\lambda_N=\int_\Lambda\langle a^*(\x)a(\0)\rangle\d\x$. This, together with the positivity of $\langle a^*(\x)a(\0)\rangle$, immediately yields that BEC is equivalent to the non-decay of $\langle a^*(\x)a(\0)\rangle$ as $|\x|\to\infty$ (after the thermodynamic limit), that is, ODLRO -- a conclusion obtained already by Penrose and Onsager \cite{rem1}.

In contrast to ODLRO which has no classical analogue, DLRO refers to classical spatial ordering of particles. The most common form of this is the periodic order of a crystal. Being classical, DLRO should be better understood than ODLRO, but this is not the case. Because of the enormous difficulty to deal with structural ordering in continuous space, we have only a few very specific rigorously proven examples of phase transitions \cite{Ru0}\,-\,\cite{Bow} and of ground-state ordering (\cite{Th}\,-\,\cite{Su2} are examples with relevance for the forthcoming discussion), all about classical systems and most of them relatively recent. A part of the new activity \cite{Su1}\,-\,\cite{Li} concerns the classical analogue of the so-called coherent crystals, a special kind of supersolids, which is the subject of the present paper. The name and a detailed study appears in Refs. \cite{KN,Ne}, but the origin is earlier, in Vlasov's theory of solids based on the Boltzmann equation \cite{Vla} and in subsequent descriptions of the ground state of fermion or boson systems in a self-consistent single-particle (Hartree or Hartree-Fock) approximation \cite{Ove}\,-\,\cite{AL}. While Vlasov's coherent crystal can be classical, that of the other papers is an intrinsically quantal object resulting from the crystallization of a quantum fluid. The conclusion of these studies is that for certain interactions at high density the expected single-particle ground state is a periodic function, therefore the particle density exhibits spatial periodicity and, hence, DLRO. For the authors of Ref.~\cite{KN} a coherent Bose crystal necessarily displays BEC, so it is a supersolid. If we disregard the approximations leading to it, the defining characteristic of a coherent crystal is that its lattice constant is not determined by the particle density but by the global properties of the interaction. As a result, when the density is increased, the lattice constant is unchanged and the number of particles within the unit cell increases.
Moreover, in the \emph{quantum} crystal the kinetic energy per particle is considerably larger than the interaction energy per pair \cite{Ne}, which can hold only if the particle mass is small and the interaction is weak. Thus, the particles move relatively freely but their motion must be coherent, otherwise they could not create and maintain an effective periodic field. The existence of coherent quantum crystals has remained hypothetical, but their classical counterpart has recently been recognized to exist in the soft matter world \cite{Li}. Below the term ``coherent crystal'' will be applied to both the classical and the quantal object. The coherent quantum crystal is understood to exhibit both DLRO and BEC. ``Coherent supersolid'' will be used as its synonym.

One can fully appreciate the peculiarities of coherent crystals only by a comparison with ``ordinary'' crystals composed of identical particles.
Ordinary crystals are thought to consist of localized particles making oscillations about separate lattice sites, their respective equilibrium position. The properties of ordinary crystals can be well described by a model based on this image, assuming harmonic (in the Born-K\'arm\'an theory) or anharmonic restoring forces \cite{BK,Pei}. Thus, each particle is tagged by a lattice site, and the pair potential does not depend on the difference of the position vectors but on the difference of the displacements from equilibrium. This means a built-in violation of permutation invariance. At a certain strength of anharmonicity the restoring forces vanish, the particles get freed and the model breaks down. If an analytic continuation beyond the singularity were possible, one could probably see the emergence of a coherent crystal.

We may say that the principal difference between coherent and ordinary crystals is revealed by their fundamentally different approximate theories outlined above. To see also the similarities, a more adequate approach would be to work in both cases with a full many-body Hamiltonian containing a permutation-invariant interaction. Then, one would never see the spontaneous breakdown of the permutational symmetry in an equilibrium state (apart from the cases of classical ground states, compressed hard-core systems or systems with unstable interactions \cite{Su0}). Only translation invariance can be broken, and the distinction becomes subtle, with a gradual transition between ordinary and coherent crystals as the interaction varies. Also, a system can show the properties of a coherent crystal at lower densities and those of an ordinary crystal at high densities. The prototype of an ordinary solid is presumably obtained with a purely repulsive interaction having a nonintegrable divergence at the origin. One then expects that the formation of a crystal at high density will be governed by short-range correlations, the lattice constant will be determined by the density and the crystal will contain a single particle in its unit cell. The prototype of a coherent crystal is provided by bounded interactions with a partly negative Fourier transform. Classically this is now confirmed \cite{Li}, and the works \cite{KN,Ne} are based on such an interaction. The repulsive square-well interaction, mentioned explicitly by Vlasov [cf. his equation (55)], has this property and defines a clear example of a classical coherent crystal \cite{Su4}.

An important intermediate case is an interaction with a strong short-range repulsion and an attractive part at a characteristic distance. Such interactions appear in Refs. \cite{Vla,Gro,AL}. The best known example is the Lennard-Jones potential, introduced to describe the Van der Waals interaction between neutral atoms and, therefore, relevant in the treatment of crystals of noble gas atoms. For He$^4$ and He$^3$ the zero-point kinetic energy can be large compared with the interaction energy of a pair, cf. \cite{AL}, but the decoupling of the lattice constant from the density may be satisfied only approximately. Because of the negative minimum in the pair potential, the interaction prefers a given structure of a well-defined density $\rho_s$, in the sense that the classical ground state energy per volume as a function of the particle density $\rho$ attains its absolute minimum on this structure. If $\rho\neq\rho_s$ but is close to it, $|\rho-\rho_s|\ll\rho_s$, without a minimum at $\rho_s$ one would expect a crystal of density $\rho$. For the interaction in question, however, the lattice density may not adjust to that of the particles but follow it with some delay, e.g., it may be $\rho_s+c(\rho-\rho_s)^3$ where $0<c<(\rho-\rho_s)^{-2}$. If this indeed occurs then the classical ground state, although still showing DLRO, is not completely ordered, there are particles or holes in excess with respect to a perfect crystal, leading possibly to a macroscopic zero point entropy. In the quantum mechanical ground state these particles or holes may condense and can be the source of ODLRO accompanying DLRO \cite{AL}. One must stress, however, that the actual knowledge about the classical ground state is insufficient to confirm the above scenario. For a two-dimensional analogue of the Lennard-Jones potential there is a proof of crystallization in the preferred structure \cite{Th} (the lattice constant is close to the distance at which the pair potential reaches its minimum), but nothing is known rigorously for $\rho\neq\rho_s$.

Some of the earliest speculations about the coexistence of ODLRO and DLRO were based less on the properties of the Hamiltonian than on the supposed form of the ground state. Taking into account the permutation (but not the translation) invariance of the Hamiltonian, the quantum mechanical ground state of a perfect ordinary crystal of identical bosons would be
\be\label{ordinary}
\Phi(\r_1,\ldots,\r_N)=C\sum_{\pi}\prod_{i=1}^N\phi\left(\r_{\pi(i)}-\R_i\right).
\ee
Here $C$ stands for normalization, the summation is over the permutations $\pi$ of $1,\ldots,N$, the product goes over the vertices $\R_i$ of a lattice in $\Lambda$, and $\phi$ is a real nonnegative $L^2$-function having a maximum at the origin and being practically independent of $\Lambda$ for large volumes.
The existence of such a ground state for a realistic Hamiltonian is as hypothetical as the existence of an ideal coherent quantum crystal whose ground state would be
\be\label{coherent}
\Psi(\r_1,\ldots,\r_N)=\prod_{i=1}^N\psi(\r_i).
\ee
Here $\psi$ is a real nonnegative, nonconstant periodic function.
It is easy to check that in the state $\Phi$ there is no ODLRO while in $\Psi$ there is. Penrose and Onsager assumed that a spatially ordered ground state can only be a perfect crystal, and naturally concluded that there can be no BEC in such a state. On the other hand, the Hartree approximation in \cite{KN} naturally lead to a coherent supersolid state of the form (\ref{coherent}). A more subtle $N$-body wave function with probable ODLRO \emph{and} DLRO was introduced by Chester \cite{Che}, whose work prompted Leggett's suggestion \cite{Leg} about the existence of a nonclassical rotational inertia in Bose solids. Apart from \cite{KN}, studies of the coexistence of ODLRO and DLRO based on the properties of the Hamiltonian are mainly restricted to lattice models. Supersolid phases, in the sense of the coexistence, were found on bipartite lattices away from half-filling by approximate analytical \cite{MT,LF} and numerical \cite{B1}\,-\,\cite{B3} methods.

This paper is meant to be a contribution to the theory of coherent quantum crystals. Even though their study has been neglected in the last years in favor of lattice models, they can play an important role in the explanation of the properties of light-atom quantum crystals at moderate pressures when the average interaction energy per pair is small compared with the kinetic energy per particle. An additional motivation comes from the recent discovery of classical coherent crystals in soft matter physics \cite{Li}. Based on earlier works \cite{KN,Ne,Li}, the appropriate interaction for the description of coherent crystals is bounded and has a partly negative Fourier transform. Without being able to deal with a full interaction, we go beyond the Hartree approximation \cite{KN,Ne} by obtaining results for a mean-field interaction which contains only Fourier components of a given wave length.

In Section 2 we recall the results on classical ground states for soft interactions and propose a method to detect the coexistence of ODLRO and DLRO in coherent quantum crystals. Section 3 contains the proof that if the Fourier transform of the interaction is partly negative, in high-density ground states of Bose systems the static structure factor develops a maximum on the support of the negative part. It is in Section 4 that we introduce and discuss the mean-field model. We prove that for a positive Fourier transform the ground state is fully Bose-condensed at all densities while if the interaction has a negative Fourier transform then at high enough densities the ground state displays joint DLRO and ODLRO. The paper ends with a few remarks in Section 5.

\section{A signature of coexisting diagonal and off-diagonal long-range orders}

In relation with the Bose gas bounded or more generally integrable pair interactions
are interesting as being at the heart of Bogoliubov's theory. Indeed, this theory is based on the Hamiltonian
\be\label{H}
H_\Lambda=\sum_{\k\in\Lambda^*}\epsilon(|\k|)N_\k+\frac{1}{2V}\sum_{\q\in\Lambda^*}v(\q)\Sigma_\q
\ee
where $\Lambda=[0,L]^d$, a $d$-dimensional cube of volume $V=L^d$ on which $H_\Lambda$ is defined with periodic boundary conditions,
$
\Lambda^*=\frac{2\pi}{L}\Zz^d
$
is the dual of $\Lambda$, $\epsilon(k)=\hbar^2k^2/2m$, $N_\k=a^*_\k a_\k$,
\be\label{Sq}
\Sigma_\q=\sum_{\k,\k'\in\Lambda^*} a^*_{\k-\q}a^*_{\k'+\q}a_\k a_{\k'}
\ee
and $a^*_\k$, $a_\k$ are the usual boson creation and annihilation operators. The underlying hypothesis in (\ref{H}) is that $v(\k)$ is the Fourier transform of an integrable pair potential $u(\r)$ and is, therefore, a bounded continuous function. An integrable $u$, even if unbounded, is quantum mechanically a soft interaction in the sense that the potential energy in a two-particle state $\Psi(\r_1,\r_2)$ may be finite even if $\Psi(\r,\r)\neq 0$. If $v$ is also integrable then $u$ is bounded and, hence, soft also classically.

Ordering of classical soft particles exhibits some peculiar features. This can already be seen in the so far unique case treated rigorously \cite{Su1,Su2} when $v(\k)\geq 0$ and identically vanishes above some wave number. Above a threshold density the classical ground state is highly degenerate, the energy-minimizing configurations are continuously deformable and their unions (even overlapping ones) are also energy-minimizing at a higher density. This is the manifestation of the striking property that any ground state configuration creates a force-free field on a test particle. According to a recent result by Likos \emph{et al.} \cite{Li}, stacking of particles in classical ground states should be even more pronounced if $v(\k)$ has a negative minimum: this minimum determines the lattice constant of a crystal, and the particles occupy the lattice points independently of the density \cite{rem0}. This seems to indicate that the crystal creates a periodic field on a test particle with \emph{minima} in lattice points, even though the pair interaction may be purely repulsive. What these authors found is, in effect, the classical coherent crystal realized in soft matter physics. It is to be noted that the results of Ref. \cite{Li} were obtained by classical density-functional theory, an analogue of the Hartree approximation used in \cite{KN}. A hint about the role of a negative minimum in $v(\k)$ is provided by the Poisson summation formula
\be\label{Poisson}
\frac{(2\pi)^{d/2}}{\sqrt{\rho_B}}\sum_{\R\in B}u(\R)=\frac{1}{\sqrt{\rho_{B^*}}}\sum_{\K\in B^*}v(\K)
\ee
written in a symmetric form. Here $B$ is a Bravais lattice of density $\rho_B$ and $B^*$ is its reciprocal lattice.
On the basis of this formula it seems reasonable that for minimizing the energy density the lattice constant of $B$ or $B^*$ should be adjusted to the minimum of $u$ or $v$, respectively. In the second case the selection of $B$ is indirect, the direct selection occurring in dual space via the dual interaction $v$.

In view of the above result on classical coherent crystals the conjecture that the ground state of the Hamiltonian (\ref{H}) with a partly negative $v$ is a coherent supersolid, showing both BEC and periodic order, seems natural. Below we argue that the coexistence of the two kinds of order can be detected solely on the off-diagonal part of the one-body reduced density matrix, without breaking translation invariance. Concerning the periodic order the suggestion is new. To mark this fact, we shall use the term SLRO (spatial or structural long-range order) instead of DLRO which can be seen on the diagonal part of the two-body density matrix. Let us recall the relevant definitions. In a translation-invariant setting the $N$-particle density matrix $\hat{\rho}_{\Lambda,N}$ ($=|\Phi_\Lambda\rangle\langle\Phi_\Lambda|$ at zero temperature, where $\Phi_\Lambda$ is the $N$-particle ground state of $H_\Lambda$) commutes with the total momentum operator and so do its reduced density matrices. With the short-hand $\langle A\rangle={\rm Tr\,}\hat\rho_{\Lambda,N} A$ 
we then obtain
\be\label{sig1}
\langle a^*(\x)a(\y)\rangle=\langle a^*(\x-\y)a(\0)\rangle=V^{-1}\sum_{\k\in\Lambda^*}\langle N_\k\rangle e^{-i\k\cdot(\x-\y)}
\ee
for the one-body reduced density matrix, and
\be\label{sig2}
\langle a^*(\x)a^*(\y)a(\x)a(\y)\rangle=V^{-2}\sum_{\k\in\Lambda^*}\langle \Sigma_\k\rangle e^{-i\k\cdot(\x-\y)}
\ee
for the diagonal part of the two-body reduced density matrix. Note that $\langle\Sigma_\k\rangle=N[S(\k)-1]$ where $S(\k)$ is the static structure factor. To detect long-range order, in the above formulas the thermodynamic limit must be taken. Below, $\lim$ without subscript will mean $V,N\to\infty$ and $N/V=\rho$. Let $\sigma^{(1)}(\x-\y)$ and $\sigma^{(2)}(\x-\y)$ be the respective limits, admitting they exist. Both are real nonnegative functions \cite{rem}, the Fourier transform of $\sigma^{(1)}$ is also positive so that $\sigma^{(1)}(\x)\leq\sigma^{(1)}(\0)=\rho$, and the average of $\sigma^{(2)}(\x)$ is $\rho^2$. Now DLRO is seen on the non-decay of $\sigma^{(2)}(\x)-\rho^2$ and ODLRO is seen on $\sigma^{(1)}(\x)\nrightarrow 0$ and on its non-vanishing average,
\be\label{avsig1}
\nu_\0=\lim_{s\to\infty}(2s)^{-d}\int_{|x_i|<s}\sigma^{(1)}(\x)\d x_1\ldots \d x_d>0,
\ee
giving the condensate density.

In the thermodynamic limit $\langle N_\k\rangle/V$ tends to some finite positive measure $\d\nu(\k)$ on $\Rr^d$. The general form of $\sigma^{(1)}(\x)$ is therefore
\be\label{measure}
\sigma^{(1)}(\x)=\int e^{-i\k\cdot\x}\d\nu(\k)=\sigma^{(1)}_{\rm ac}(\x)+\sigma^{(1)}_{\rm pp}(\x)+\sigma^{(1)}_{\rm sc}(\x),
\ee
the three terms corresponding to the decomposition of $\d\nu(\k)$ into absolutely continuous (ac), pure point (pp) and singularly continuous (sc) parts.  For (\ref{measure}) to hold, $\int\d\nu(\k)=\rho$ must be satisfied, thus,
\be
F(k_0)=\int_{|\k|<k_0}\d\nu(\k)=\lim\sum_{\k\in\Lambda^*,|\k|<k_0}\frac{\langle N_\k\rangle}{V}\leq\rho
\ee
must tend to $\rho$ as $k_0\to\infty$. This is indeed the case, otherwise the kinetic energy density would diverge in the thermodynamic limit because of the inequality
\be\label{tdiv}
t\equiv\lim\sum_{\k\in\Lambda^*}\epsilon(|\k|)\frac{\langle N_\k\rangle}{V}
\geq\epsilon(k_0)\left[\rho-F(k_0)\right].
\ee
Actually, $F(k_0)\geq\rho-c/k_0^2$ with some $c>0$ as $k_0\to\infty$. 

Due to the Riemann-Lebesgue lemma, $\sigma^{(1)}_{\rm ac}(\x)$ tends to zero as $|\x|$ goes to infinity. This is not true for $\sigma^{(1)}_{\rm pp}(\x)$ and $\sigma^{(1)}_{\rm sc}(\x)$. The pp and sc parts of $\d\nu$ are concentrated on sets of zero Lebesgue measure and can be interpreted as different kinds of (generalized) BEC \cite{BL}. However, the pure point part
\be\label{nusum}
\sigma^{(1)}_{\rm pp}(\x)=\nu_\0+\sum_{\k\in D^{(1)}}\nu_\k e^{-i\k\cdot\x}
\ee
where $D^{(1)}$ is a countable subset of $\Rr^d$ not containing $\0$ has another, more relevant interpretation. In \cite{PVZ} it was shown that if $\nu_\k>0$ for some $\k\neq \0$ then there exist infinite-volume equilibrium states which are periodic in the direction $\k$. More generally, the sum in Eq.~(\ref{nusum}) can describe almost periodic order if $D^{(1)}$ is a general nonempty set, and $B$-periodic order if $D^{(1)}=B^*$. Whether or not $\sigma^{(1)}_{\rm sc}(\x)$ can be nonzero is unknown, but in principle it can also contribute to the asymptotic ($|\x|\to\infty$) non-constancy of $\sigma^{(1)}(\x)$. In the definition below we disregard this possibility.

\textbf{Definition.} We say that a density matrix displays spatial or structural long-range order (SLRO) if its associated set $D^{(1)}$ defined in Eq.~(\ref{nusum}) is nonempty.

From the definition it follows that SLRO implies BEC. Indeed, if $\nu_\k>0$ for some $\k\neq \0$ then $\nu_\0>0$ as well:
\be
\nu_\k=\lim_{s\to\infty}(2s)^{-d}\int_{|x_i|<s}e^{i\k\cdot\x}\sigma^{(1)}(\x)\d x_1\ldots \d x_d\leq\nu_\0
\ee
because $\sigma^{(1)}\geq 0$. Furthermore, SLRO implies also DLRO. This is an interpretation of the result of Ref. \cite{PVZ} and follows from the meaning of $\nu_\k=\nu_{-\k}>0$: If translation invariance were broken e.g. by an arbitrarily weak $\k$-periodic external field, one would see the macroscopic occupation of a one-particle state $\sim\cos(\k\cdot\r+\alpha)$. This would imply the spatial periodicity of the density and would show up in the diagonal part of the one-body density matrix. Hence, we conclude that SLRO implies the coexistence of ODLRO and DLRO. 

\section{Effect of a partly negative $v(\k)$ on the ground state of a system of bosons}

In this short section we show that whenever the Fourier transform of the interaction has a negative part $v_-$, at high enough densities the static structure factor $S(\k)$ develops a maximum on the support of $v_-$. Decompose $v$ into positive and negative parts,
\be\label{vpm}
v(\k)=v_+(\k)-v_-(\k)
\ee
where $v_\pm\geq 0$ and $v_+v_-= 0$, and suppose that $v_-$ is not identically vanishing. Because $u\in L^1(\Rr^d)$, $v$ is a bounded continuous function decaying at infinity, therefore $v(\k)<0$ on a union of open intervals. Choose a $\k_0$ such that $v(\k_0)<0$, and compute the energy in the product state
\be\label{Psi}
\Psi(\r)_N=\prod_{j=1}^N\sqrt{\frac{2}{V}}\cos\frac{1}{2}\k_0\cdot\r_j.
\ee
Here $(\r)_N=(\r_1,\ldots,\r_N)$. For
\be\label{tildeH}
\tilde{H}_\Lambda=H_\Lambda-\frac{v(0)}{2V}N(N-1)
\ee
by a simple computation one finds
\be
\frac{1}{V}\langle\Psi|\tilde{H}_\Lambda|\Psi\rangle=\frac{1}{4}[\epsilon(|\k_0|)+\rho v(\k_0)]\rho.
\ee
This is an upper bound to the ground state energy density,
\be\label{boundSk}
\frac{1}{V}\langle\Phi_\Lambda|\tilde{H}_\Lambda|\Phi_\Lambda\rangle=t+\frac{\rho}{2V}\sum_{0\neq\k\in\Lambda^*}v(\k)[S(\k)-1]\leq \frac{1}{4}[\epsilon(|\k_0|)+\rho v(\k_0)]\rho.
\ee
Insert here (\ref{vpm}), use the lower bound 0 for $S(\k)$ on ${\rm supp\,} v_+$ and also for $t$ to find
\be\label{weight}
\frac{\sum_{\k\in\Lambda^*}v_-(\k)[S(\k)-1]}{\sum_{\k\in\Lambda^*}v_-(\k)}\geq\frac{\rho|v(\k_0)|-\epsilon(\k_0)
-\frac{2}{V}\sum_{0\neq\k\in\Lambda^*}v_+(\k)}{\frac{2}{V}\sum_{\k\in\Lambda^*}v_-(\k)}.
\ee
Note that $v(\0)=\int u(\r)\d\r>0$ and therefore $v(\k)>0$ in a neighborhood of $\k=\0$. The weighted mean of $S(\k)-1$ on the left-hand side of (\ref{weight}) becomes positive if $\rho$ is above a volume-independent threshold value and then it increases at least linearly with $\rho$. Since the arithmetic mean of $S(\k)$ over $\Lambda^*\setminus\{\0\}$ is 1, we can indeed see the appearance of a maximum of $S(\k)$ somewhere on ${\rm supp\,} v_-$. This still does not prove DLRO | for that we should show that $S(\k)\sim N$ for some $\k\neq\0$ at a fixed density $\rho$ as $V,N\to\infty$, $N/V=\rho$ | but indicates its possible occurrence.

\section{A mean-field model for coherent supersolids}

Unfortunately, the present mathematical knowledge does not allow one to prove even BEC alone for the Hamiltonian (\ref{H}) with a general $v(\k)$. However, we can check the ideas outlined in Section 2, in particular the role of a negative minimum of $v(\k)$ in a simplified case. Let us recall that
\be\label{v-u}
v(\k)=\int_{\Rr^d}u(\r)e^{-i\k\cdot\r}\d\r=\int_{\Lambda}u_\Lambda(\r)e^{-i\k\cdot\r}\d\r,
\ee
where
\be
u_\Lambda(\r)=\sum_{\n\in\Zz^d}u(\r+L\n)
\ee
is the periodized interaction that we normally use when we work with periodic boundary conditions. By inverting (\ref{v-u}),
\be\label{exp}
u_\Lambda(\r)=\frac{1}{V}\sum_{\k\in\Lambda^*}v(\k)e^{i\k\cdot\r}.
\ee
The constant first term, $v(\0)/V$, is known as the mean-field interaction for Bose gases. Although in the canonical ensemble the corresponding mean-field model does not differ from the ideal Bose gas, the mean-field interaction is superstable (the potential energy of $N$ particles is $v(\0)N(N-1)/2V$), which makes it possible to treat the system in the grand-canonical ensemble for arbitrarily large values of the chemical potential and, hence, in the region of BEC. The mean-field model has been studied extensively in the past; rigorous results can be found e.g. in Refs.~\cite{Dav}-\cite{BP}. The mean-field interaction is interesting also as the limiting case of a scaled interaction (the Kac potential); this observation was the basis of recent attempts \cite{MP1,MP2} to prove BEC for nontrivial interactions. If we go further in the expansion (\ref{exp}) by retaining a finite number of terms, the result is still a mean-field interaction which is, however, not constant. By a proper choice of the additional terms one may hope to preserve at least some qualitative features of the fully interacting system. To demonstrate that a negative minimum in $v(\k)$ may lead to joint DLRO and ODLRO the proper choice is the set of wave vectors of length $q$ at which the minimum is attained. For a cubic domain this leads to the interaction
\be\label{u}
u_\Lambda(\x)=\frac{a}{V}+\frac{2b}{V}\sum_{j=1}^d\cos\q_j\cdot\x,
\ee
where $\q_j=q\e_j$. Let $L$ be an integer multiple of $2\pi/q$, then $\Lambda=[0,L]^d$ is a period cell for the Bravais lattice $B=(2\pi/q)\Zz^d$, and $\Lambda^*=(2\pi/L)\Zz^d\supset B^*=q\Zz^d$. The simple cubic $B$ is just an example. With suitable $\q_j$ and $\Lambda$, $B$ can be any Bravais lattice. If $a>2d|b|$ then $u_\Lambda$ is superstable. However, $a$ does not affect the eigenstates of $H_\Lambda$, and the potential energy density at a fixed particle density is bounded below without the term $a/V$. We can therefore set $a=0$, and for comparison we shall consider both $b>0$ and $b<0$. As the theorem below will show, for $b<0$ (\ref{u}) defines the mean-field model of coherent supersolids in continuous space.

Before proceeding further, let us comment on the possible relevance of the mean-field model in the study of the Hamiltonian (\ref{H}). 
A mean-field-type truncation (\ref{u}) of a soft
interaction is a high-density approximation.
If $v\geq 0$, there is a clear indication for this being true. In effect, from Eqs.~(\ref{tildeH}) and (\ref{boundSk}), and by using
$\Omega(\r)_N\equiv V^{-N/2}$ as a variational wave function instead of (\ref{Psi}),
one obtains the bounds
\be
\frac{v(0)}{2}\rho^2-\frac{u(0)}{2}\rho\leq e_{\rm g.s.}=\lim\frac{1}{V}\langle\Phi_\Lambda|H_\Lambda|\Phi_\Lambda\rangle\leq \frac{v(0)}{2}\rho^2
\ee
for the ground state energy density (see also the Appendix of Lieb \cite{Lie}). Thus, as $\rho$ tends to infinity, the leading term in $e_{\rm g.s.}$ is $v(0)\rho^2/2$, the exact result for the mean-field Bose gas.
If $v$ is partly negative, the precise asymptotic ($\rho\to\infty$) form of the ground state energy density is unknown. It is clear, however, that minimization of the potential energy dictates that if the classical ground state is ordered, this order shows up in the quantum mechanical ground state. Our discussion in the previous section supports this expectation. The result obtained by Likos et al. \cite{Li,rem0}
reveals the order characteristic to classical coherent crystals, particles accumulating on the sites of a given lattice in high-density ground states. It will be seen that the mean-field interaction (\ref{u}) has such ground states for $b<0$. Let us note that in Ref.~\cite{KN} explicit mention is made of a pair potential of the form (\ref{u}) as an interaction that in Hartree approximation would lead to a ground state in which the only possible momentum transfer is for $|\k|=q$. For the full many-body ground state this view will prove to be over-simplified.
Finally, we stress that although with (\ref{u}) a periodicity is introduced by hand, the interaction is still translation-invariant, a case not to be confounded with the presence of a periodic external field as in optical lattices \cite{ALS}. Moreover, the periodicity will be seen to disappear in the thermodynamic limit if $b>0$ and to stay if $b<0$ \emph{and} the density is high enough. The difference between the two cases is the main point of the following theorem.

\textbf{Theorem.} For $b>0$ the ground state of $H_\Lambda$ is asymptotically fully Bose-condensed at any $\rho$. For $b<0$ and $\rho>(e/4)\epsilon(q)/|b|$ the ground state presents both BEC and SLRO/DLRO in the sense that both $\sigma^{(1)}$ and $\sigma^{(2)}$ are nonconstant $B$-periodic functions.

\emph{Proof.} 1. We start with identifying the classical ground state configurations and energies. The $N$-particle interaction energy is
\be\label{pot}
U_\Lambda(\r)_N=\frac{1}{2}\sum_{i\neq j}u_\Lambda(\r_i-\r_j)=
\frac{b}{V}\sum_{j=1}^d\left|\sum_{n=1}^Ne^{i\q_j\cdot\r_n}\right|^2-\frac{dbN}{V}.
\ee
For $b>0$,
\be\label{bge0}
U_\Lambda(\r)_N\geq-dbN/V
\ee
and the lower bound is actually the classical ground state energy. Any $(\r)_N$ making $\sum_{n=1}^Ne^{i\q_j\cdot\r_n}$ disappear for all $j$ is a classical ground state. Each of these $(\r)_N$ can be obtained by the following construction. Writing $\r_n=(2\pi/q)(x_{1n},\ldots,x_{dn})$, $\q_j\cdot\r_n=2\pi x_{jn}$. For each $j$ we partition the coordinates $\{x_{jn}\}_{n=1}^N$ into groups of at least 2 elements and in a group of $m$ elements we choose a real $\alpha$ and assign the values $\alpha,\alpha+1/m,\ldots,\alpha+(m-1)/m$ to the coordinates in an arbitrary order. The sum of the exponentials will disappear separately for each group. In this highly degenerate set of ground state configurations only a negligible number shows long-range order. All will be mixed in a unique unordered quantum mechanical ground state; that this one is asymptotically fully Bose-condensed, is not \emph{a priori} obvious.

For $b<0$,
\be\label{ble0}
U_\Lambda(\r)_N\geq-d|b|N^2/V+d|b|N/V
\ee
and the lower bound is again the classical ground state energy. All $(\r)_N$ giving $e^{i\q_j\cdot\r_n}=1$ for each $j$ and $n$ and all translates of such configurations are classical ground states. Independently of $N$, these are the $(\r)_N$ with $\r_n-\r_m\in B$ for every $m,n$. Thus, for $b<0$ in any classical ground state configuration all particles occupy the points of a lattice whose lattice constant is independent of the particle density. Classically, the simple interaction (\ref{u}) reproduces the zero-temperature result of Likos \emph{et al.} \cite{Li,rem0}.

2. The Hamiltonian (\ref{H}) with
\be\label{int}
v(\k)=a\delta_{\k,\0}+b\sum_{j=1}^d(\delta_{\k,\q_j}+\delta_{\k,-\q_j})
\ee
corresponding to (\ref{u}) has a large number of extra symmetries. Consider the coset group $\Lambda^*/B^*$ of $\Lambda^*$. For any $C\in\Lambda^*/B^*$, $H_\Lambda$ commutes with $N_C=\sum_{\k\in C}N_\k$. Consider now any partition of $N$, $1\leq n_1\leq\cdots\leq n_p\leq N$, $\sum n_i=N$, and any $C_1,\ldots,C_p\in \Lambda^*/B^*$. Then the vectors $\Psi$ satisfying $N_{C_j}\Psi=n_j\Psi$ for $j=1,\ldots,p$ form an $H_\Lambda$-invariant subspace of $[L^2(\Lambda)^N]_{\rm sym}$. For the $N$-particle ground state $\Phi_\Lambda$ of $H_\Lambda$, $N_{B^*}\Phi_\Lambda=N\Phi_\Lambda$, meaning that $\Phi_\Lambda$ is constructed solely from plane waves whose $\k$ vector is in $B^*$. Indeed, $\Phi_\Lambda$ is the eigenstate of $e^{-\beta H_\Lambda}$ belonging to the largest $N$-particle eigenvalue. (Here $\beta$ is an arbitrary positive number of dimension 1/energy.) Because the integral kernel of $e^{-\beta H_\Lambda}$ is positive, $\Phi_\Lambda$ is positive and hence unique. Therefore, it must show all symmetries of $H_\Lambda$. Thus, $\Phi_\Lambda$ is permutation and translation invariant, and $B$-periodic separately in each position vector. The only invariant subspace whose elements have these symmetries is $N_{B^*}\Psi=N\Psi$.

3. In the ground state, $\langle\cdot\rangle=\langle\Phi_\Lambda|\cdot|\Phi_\Lambda\rangle$ and (\ref{sig1}) reads
\be\label{sig1gs}
\langle a^*(\x)a(\y)\rangle=
N\int_{\Lambda^{N-1}}\Phi_\Lambda\left(\x,(\r)_2^N\right)\Phi_\Lambda\left(\y,(\r)_2^N\right)\d(\r)_2^N.
\ee
According to point 2,
\be\label{sig11}
\langle a^*(\x)a(\y)\rangle=V^{-1}\sum_{\k\in B^*}\langle N_\k\rangle e^{-i\k\cdot(\x-\y)}.
\ee
So $\langle a^*(\x)a(\0)\rangle$ is $B$-periodic in finite volumes. By the diagonal process we can choose a subsequence of $L$ values such that $\langle N_\k\rangle/V$ converges to some $\nu_\k$ for every $\k\in B^*$ as $V$ and $N=\lfloor\rho V\rfloor$ tend to infinity. We have then three possibilities. (i) $\sigma^{(1)}(\x)\equiv\rho=\nu_\0$, meaning full BEC. (ii) $\sum_{\k\in B^*}\nu_\k=\rho>\nu_\0$.
Because $\nu_\k\leq \nu_\0$, there is partial BEC ($\nu_\0>0$) and partial SLRO ($\nu_\k>0$ for some nonzero $\k\in B^*$), realized as a sort of full generalized BEC, $\sigma^{(1)}(\x)=\sigma^{(1)}_{\rm pp}(\x)$ is a nonconstant $B$-periodic function. (iii) $0\leq\sum_{\k\in B^*}\nu_\k<\rho$, which can immediately be excluded because it would lead to a diverging kinetic (and total ground state) energy density, cf. Eq.~(\ref{tdiv}). In cases (i) and (ii) the infinite sum and the limit are interchangeable in $\lim\sum_{\k\in B^*}\langle N_\k\rangle/V(=\rho)$.

4. Let $b>0$. Using Eqs.~(\ref{pot}),~(\ref{bge0}) and $\Omega(\r)_N\equiv V^{-N/2}$ as a variational wave function, we obtain
\be
-dbN/V=\min_{(\r)_N} U_\Lambda(\r)_N\leq\langle\Phi_\Lambda|U_\Lambda|\Phi_\Lambda\rangle
\leq\langle\Phi_\Lambda|H_\Lambda|\Phi_\Lambda\rangle\equiv E_{\rm g.s.}\leq\langle\Omega|H_\Lambda|\Omega\rangle
=0.
\ee
The last equality holds because both the kinetic and the potential energies vanish in the state $\Omega$. Dividing by $V$ and taking the thermodynamic limit this yields $e_{\rm g.s.}=\lim\langle\Phi_\Lambda|H_\Lambda|\Phi_\Lambda\rangle/V=0$, and the kinetic and potential energy densities separately vanish in the ground state. On the other hand,
\be\label{tlow}
t=\sum_{\k\in B^*}\epsilon(|\k|)\nu_\k\geq\epsilon(q)\sum_{\0\neq\k\in B^*}\nu_\k=\epsilon(q)(\rho-\nu_\0).
\ee
If $\nu_\0<\rho$, we would obtain $t>0$; thus, $\nu_\0=\rho$ which is case (i) of point 3. This proves the $b>0$ part of the theorem.

5. Let $b<0$. First we show that $\sigma^{(2)}$ is a nonconstant periodic function if the density is high enough. The ground state energy density reads
\be
e_{\rm g.s.}=t-|b|d\rho s_q
\ee
where
\be
s_q=\lim S(\q_j)/V\quad(\mbox{for any $j$}).
\ee
With the variational ansatz $\Omega$ and Eq.~(\ref{ble0}) we find $-d|b|\rho^2\leq e_{\rm g.s.}\leq0$. A better upper bound is obtained with the trial function
\be
\Psi(\r)_N=\prod_{n=1}^N\psi(\r_n),\quad \psi(\r)=(q^2/2\pi)^{d/4}e^{-q^2 r^2/4}
\ee
where the parameter of the Gaussian is already optimized. This yields
\be\label{eup}
e_{\rm g.s.}\leq d\rho[\epsilon(q)/4-|b|\rho/e]
\ee
which is negative if $\rho>(e/4)\epsilon(q)/|b|$. Thus, $s_q>0$ for $\rho>(e/4)\epsilon(q)/|b|$, and the expansion (\ref{sig2}) (which also reduces to $B^*$) reads
\be
\sigma^{(2)}(\x)=\rho^2+2\rho s_q \sum_{j=1}^d \cos\q_j\cdot\x+\cdots
\ee

To prove the periodicity of $\sigma^{(1)}$ we start with Eq.~(\ref{Sq}), apply the Schwarz inequality to the addends, the Cauchy inequality to the sum and the equality $\sum_{\k\in\Lambda^*}N_\k=N\times\,{\rm Id}$. The result is
\be\label{Squp}
S(\k)\leq 1+N-\left(\sqrt{\langle N_\0\rangle}-\sqrt{\langle N_\k\rangle}\right)^2,
\ee
holding in all generality. Dividing by $V$ and letting $N$ and $V$ tend to infinity, for $\k=\q_j$ at $\rho>(e/4)\epsilon(q)/|b|$ we find
\be\label{ii}
\left(\sqrt{\nu_\0}-\sqrt{\nu_{\q_j}}\right)^2\leq\rho-s_q<\rho.
\ee
Now either $\nu_\0\geq\nu_{\q_j}>0$ or $\nu_{\q_j}=0$ and then $\nu_\0<\rho$ follows from (\ref{ii}). Because $\sum_{\k\in B^*}\nu_\k=\rho$, $\nu_\0\geq\nu_\k >0$ for some nonzero $\k\in B^*$. This ends the proof of the theorem.

\section{Concluding remarks}

1. We have not proven that for $b<0$ there is no DLRO at low densities, but based on Eq.~(\ref{eup}) the existence of a $\rho_c\sim\epsilon(q)/|b|$ seems plausible. The order of the quantum phase transition at $\rho_c$ is an interesting open question. In Refs.~\cite{KN,Ne} there is a discussion of this point. It is argued that whenever there exists a triplet of reciprocal lattice vectors such that $|\k_i|=q$ and $\k_1+\k_2+\k_3=\0$ then by Landau's argument coherent crystallization is a first order transition. This would be the case if, instead of a simple cubic $B$, we would choose $B$ to be the basic-centered cubic lattice or the hexagonal or rhombohedral lattice with some specific aspect ratios. On the other hand, if no such triplet exists -- this is the case e.g. of a simple cubic $B$ -- then, because the interaction (\ref{u}) is anisotropic, the transition could be of second order, cf. \cite{Ne}.

\noindent
2. It may seem curious that DLRO is proven in a translation invariant situation, therefore the state in infinite volume is not a pure one but a uniform mixture over translations. However, the same holds true for the proof of BEC: by not breaking gauge invariance, the infinite-volume state is a uniform mixture over the phases of condensates \cite{FPV}\,-\,\cite{Su3}.

\noindent
3. In our proof we have not used that SLRO implies DLRO but proved the periodicity of $\sigma^{(1)}$ and $\sigma^{(2)}$ separately. It should be possible to prove the above implication in the general case without referring to the breakdown of translation invariance, by showing that the asymptotic periodicity of $\sigma^{(1)}$ implies that of $\sigma^{(2)}$.

\noindent
4. The way $\Phi_\Lambda(\r)_N$ describes $B$-periodic order is somewhat special, since it is $B$-periodic separately in each $\r_n$.
In general for a translation-invariant ground state one can expect only some weaker property, e.g. it having maxima at configurations $(\r)_N$ such that $\r_n-\r_m\in B$ for a positive fraction, nonvanishing in the limit of infinite volume, of pairs of particles (and $\r_m=\r_n$ excluded  for interactions that are non-integrable at the origin).

\noindent
5. The possible appearance of SLRO in the one-body reduced density matrix for a general soft interaction is better seen on the ground state formula (\ref{sig1gs}). Suppose that $\Phi_\Lambda$ describes $B$-periodic order in the weak sense given above. If $\x-\y\in B$, $(\r)_2^N$ can be chosen in many different ways so that both $\Phi_\Lambda\left(\x,(\r)_2^N\right)$ and $\Phi_\Lambda\left(\y,(\r)_2^N\right)$ are at a maximum. Then the integral itself may be at a maximum.

\noindent
6. Although the Poisson duality (\ref{Poisson}) places $u$ and $v$ on equal footing, quantum mechanically this symmetry is broken by the kinetic energy term. This is the reason why we expect in general different results when $u$ has negative minima but $v$ is positive and when $v$ has negative minima and $u\geq 0$, even though in both cases there can be a lattice of a specific density minimizing the potential energy. 

\noindent
7. The mean-field result suggests that in the ground state of the Hamiltonian (\ref{H}) there is no DLRO without BEC. This may not be the case at positive temperatures. A crystal may be formed at a temperature higher than the critical temperature of BEC at the given density. If now the temperature is kept fixed and the density is increased, the systems responds as a classical coherent crystal, by the accumulation of particles in each unit cell. However, increasing the density at a given temperature leads also to BEC: in this respect a system with a bounded interaction should not differ from the ideal Bose gas. Passing the critical density for BEC marks the transition from classical to quantum coherent crystal.
\vspace{3pt}

\noindent
\textbf{Acknowledgement}. I thank the referee for drawing my attention to some publications, in particular to references \cite{KN} and \cite{Ne}. This work was supported by OTKA grants T 46129 and 67980K.


\begin{thebibliography}{99}
\bibitem{rem}
The one- and two-body density matrices derive from the integral kernel of $e^{-\beta H_\Lambda}$ whose positivity is seen via the Feyman-Kac formula.
\end{thebibliography}
\end{document}